# ALGEBRAIC QUANTUM MECHANICS AND PREGEOMETRY

## BY

D.J. Bohm, P.G. Davies and B.J. Hiley

Birkbeck College (University of London) Malet Street

London WC1E 7HX

1981

**Abstract**

*We discuss the relation between the q-number approach to quantum mechanics suggested by Dirac and the notion of 'pregeometry' introduced by Wheeler. By associating the q-numbers with the elements of an algebra and regarding the primitive idempotents as 'generalized points' we suggest an approach that may make it possible to dispense with an a priori given space manifold. In this approach the algebra itself would carry the symmetries of translation, rotation, etc. Our suggestion is illustrated in a preliminary way by using a particular generalized Clifford Algebra proposed originally by Weyl, which approaches the ordinary Heisenberg algebra in a suitable limit. We thus obtain a certain insight into how quantum mechanics may be regarded as a purely algebraic theory, provided that we further introduce a new set of "neighbourhood operators", which remove an important kind of arbitrariness that has thus far been present in the attempt to treat quantum mechanics solely in terms of a Heisenberg algebra.*

## 1. Introduction

Quantum field theories are generally constructed by assuming a basic a priori given space-time structure where the coordinates are treated as $x^\mu$ are treated as parameters belonging to a pre-Hilbert space. The success of this formalism particularly for the electromagnetic and weak forces suggest that a similar method should be carried through for the quantization of gravity where the metric tensor now becomes the subject of quantization. Unfortunately renormalisation presents serious difficulties to such a programme and one of the more recent suggestions is that the root of the problem may lie in the intimate relation between gravity and the space-time structure itself. There is now a growing realisation that perhaps the use of a differential manifold will have to be called into question and other possibilities explored (see, for example, Taylor, 1979, t'Hoof, 1978 and Wheeler 1980).

Indeed a decade ago Finkelstein (1972) had already expressed dissatisfaction with quantum field theory by pointing out that present theories are essentially hybrids in which classical space-time (c) is combined with quantum matter (q). What is required, he suggests, is not a cq-theory but a purely q-theory where no reference is made to an a priory given space-time. In such an approach, space and time would emerge from some deeper theory. Since the deeper theory can no longer use the properties of a differential manifold in a basic way we should follow Wheeler (1980) and regard such a theory as 'pregeometric'. But this immediately raises the question as to the nature of the elements of such a theory. The purpose of this paper is to examine this question in the context of the present quantum formalism which we will analyse in a manner that is basically different from the usual one. We will show that our approach leads to a possibility of realising such a theory although at this stage we will make no attempt to connect our structure with gravity.

Our ideas start by noting that a 'pregeometric' theory is already to some extent implicit in the original Heisenberg matrix approach to quantum theory when what is regarded as 'position' becomes part of a matrix algebra. Here the 'position' was represented as a matrix q(n,m) which, in the way it was developed, appeared to be more like a transition or 'two-point' object than a means of labelling the points of a continuum. It was only through the Schrödinger representation that $|X\rangle$ became associated with a way of labelling points of an underlying differential manifold. The operator $X$ is taken as a description of the way we locate objects at some position in space using a suitable measuring instrument. The fact that the two approaches (Heisenberg and Schrödinger) have generally been regarded as equivalent has led to the main emphasis being placed on the Schrödinger representation which, as is well known, has what appears to be a particularly simple interpretation through the wave function. The success of the formalism has strongly reinforced the idea that the underlying differential manifold must play a fundamental role in the theory.

When we consider the case of a system with an infinite number of degrees of freedom we find that the above equivalence breaks down. Dirac (1963) has illustrated this point by means of a simple example in which he shows that Heisenberg's approach yields results that cannot be derived from a wave function even though the energies are well defined. He also points out that some of the infinites of the conventional theories are no longer present in the q-number approach.

As a result of this work, Dirac suggests that perhaps the q-number approach should be taken more seriously, but one of the main problems in developing such a theory lies in the apparent lack of any compelling physical interpretation of the q-numbers. Recently, Bohm (1973) has developed some very general ideas involving what is called the implicate order. This approach, we believe, begins to provide a more general framework in which to view q-number formalism, and although we will occasionally by guided by general considerations appearing in the implicate order, much of our work stems directly from the mathematical structures used in this paper.

We will not discuss the case of a system with an infinite number of degrees of freedom. Rather, our aim here will be to motivate an exploration of q-number theories and to do this we will consider the finite case the only[1]. Furthermore we will restrict ourselves at this stage to any discussion of the non-relativistic theory in order to make clear the concepts used in these initial investigations.

Some preliminary work has already been done by Frescura and Hiley (1980a). They indicate in detail some general lines along which one can make an essentially algebraic approach to quantum mechanics. What they do is to represent <u>all</u> physical features by the elements (q-numbers) of some suitable algebra, the nature of which is determined by the physical context. In such an approach there is no need for the disjoint features of the present mathematical formalism, namely, the operators on the one hand and the state space vectors on the other. Rather, one uses only a single type of object, the algebraic element (q-number). What is now taken to be the state factor is simply a distinguished element in the algebra, namely, the minimal ideal.

This idea is most easily illustrated in the case of the Pauli and the Dirac-Clifford algebras and, indeed, such as a possibility was already anticipated by Riesz (1946). The polynomial Heisenberg algebra studied by Born and Jordan (1925) appeared to have no minimal ideals (except the trivial ones) so that the generalization of Reisz's work to quantum mechanics seemed impossible. However, a new element can be introduced into the polynomial algebra from which the minimal ideals corresponding to the state vectors can be generated. This new element plays a role similar to the standard ket in Dirac's (1947) bra - ket notation (Frescura and Hiley, 1980b). It was this additional feature that allows for the possibility of a completely q-number approach. In this paper we will try to bring out more clearly how the q-number theory is related to the usual bra-ket notation and, as a consequence, we will show that the q-number theory is potentially richer even in the case of a finite number of degrees of freedom. Furthermore we will bring out how the q-number approach does not require an a priori given space manifold. In this sense it can be regarded as 'pre-geometric'.

In section 2 we will show how such a pregeometric structure can be given an algebraic meaning in terms of the idempotents and their corresponding minimal left and right ideals. These features provide some of the necessary concepts that are needed to understand the q-number approach in which space is not taken as an a priori given structure. In the section 3, we discuss how these concepts can be illustrated through a particular algebraic structure namely the collection of algebras known as the generalized Clifford algebras (Morris, 1967). These algebras have been studied in some detail by Ramakrishnan, Santhanam and their coworkers (see Santhanam 1977, for an extensive list of references)

We will discuss a particular generalized Clifford algebra $C^n_2$ which can be used to set up quantum mechanics in a discreet one dimensional space. In the limit as n → ∞, $C^n_2$ approaches the Heisenberg algebra as was pointed out many years ago by Weyl (1932) to. Santhanam and Tekumella (1975) have already used this structure to suggest a discrete quantum mechanics using the vector space approach. We analyze this structure If purely from the q-number point of view and show in section 4 that contrary to the conclusions of Santhanam and Tekumella (1975) the discrete space does possess an uncertainty principle.

In the final section we discuss how the geometric order emerges from the pregeometric algebraic structure. In particular we find that the physical meaning of the Heisenberg algebra can be determined only by introducing a new set of neighbourhood operators, which are parts of the algebra in this way we remove a certain arbitrariness that is inherent in the very concept of a Heisenberg representation, and determine which algebraic elements are the ones that represent the actual physical space.

**2. The Role of the Primitive Idempotent**

In order to motivate our approach, let us briefly recall some of the ideas that prompted Wheeler to suggest some form of pregeometric theory[2]. In the classical approach space-time is assumed to be a continuum. In order to give meaning to matter and gravity in this context, Sakharov (1967) has suggested the possibility of regarding the continuum has some form of 'elastic' medium out of which matter is formed. The concentrations of matter then leave something analogous to a stress in the medium. The average properties of these stresses can then be described by the curvature tensor which in turn can be associated with the average distribution of matter so that a petition becomes the 'metric elasticity' of space.

However in the case of ordinary matter, elasticity has its origins in the atoms and the forces between them; atoms are not built from elasticity. Could not a metric geometry be hiding a deeper 'atomic' structure which is revealed in some way by the appearance of particle-anti-particle patterns which arise from fluctuations of the vacuum. By associating charge with small-scale topological features of space-time, Wheeler argues that the quantum fluctuations with their implications for a continual change in topology suggests something more like a foam (Hawking 1976), or perhaps like some form of disordered lattice structure continually undergoing restructuring (Hiley 1980). For such structures the continuum with its well defined and fixed neighbourhood relations may not be an appropriate starting point. Rather we should begin with a set of basic elements, which, for convenience, we can call 'generalized points', and the relationships between them. This will give rise to a generalized structure in which neither a fixed neighbourhood relation nor a fixed dimensionality have any direct relevance in the small-scale. Our conventional space-time is then an abstraction which emerges from this structure through some form of macroscopic averaging.

Wheeler's specific suggestion was to build the pregeometry from a set of Boolean elements giving a dichotomic choice such as yes or no, true or false, on or off, etc. Although his original hope was to use some of the notions of formal logic to lay the foundations of pregeometry, he has since concluded that, in fact, formal logical leads in another direction away from quantum theory rather than towards it. However his conclusion should not be taken to imply that the simple dichotomic element cannot be used as the basic descriptive form in pregeometry. What we will do here is to draw attention to the fact that such a elements play an important role in the analysis of algebras in general and, furthermore, the dichotomy of yes/no arises in an essentially quantum mechanical way, namely through the eigenvalues of the set basic elements. Here we are referring to the primitive idempotents.

This idea has already been discussed briefly in Frescura and Hiley (1980a). They argued that the link between the usual geometric entities, vectors, bivectors, etc., appeared a ordered products of the elements taken from minimal left and right ideals. But these minimal ideals are generated from a set of primitive idempotents $\varepsilon_i$ which satisfies the relations

$$\varepsilon_i^2 = \varepsilon_i$$
$$\varepsilon_i \varepsilon_j = \varepsilon_j \varepsilon_i = 0 \qquad (1)$$
$$\sum_i \varepsilon_i = 1$$

These idempotents have If eigenvalues 1 and 0 so that they become the algebraic equivalents to the yes-no logic.

It is clear that idempotents can be constructed in quantum mechanics where they are, of course, called projections operators. These projection operators are used to form the basis of the propositional calculus first introduced by Berkoff and von Neumann (1936) which has since been developed into a formal structure called quantum logic (see, for example the Jauch 1968). We do not want to follow this approach for reasons similar to those given by Wheeler (1980) and, more importantly, because here we want to open up new possibilities, the relevance of which cannot be seen if the idempotents are treated as propositions. Indeed there is no notion of projections in our algebraic structure.

Instead we will follow Eddington (1946) who argued that within a purely algebraic approach, which he regarded as providing a structural description of physics, there are elements of existence defined, not in terms of some hazy metaphysical concept of existence, but in the sense that existence is represented by a symbol which contains only two possibilities - existence or non-existence. Thus we will assume that the structural concept of existence is represented by an idempotent of some appropriate algebra. But recalling that any idempotent can always be decomposed into primitive idempotents satisfying the relations (1), we will take the primitive idempotents as our basic descriptive forms. They will be considered as the generalized points of our structure.

It should be noted that these primitive idempotents, although being used as basic elements of the description, should not be considered as some form of absolute element of reality. The symbols do not have any direct meaning in isolation. They are not dinge-an-sich but take their meanings from within the overall context of the given algebra, which, in turn, is determined from a particular physical context. In this way, we incorporate Bohr's notion of 'wholeness' and d'Espagnat's view of 'non separability' in a very basic way. In this sense our generalize points would appropriately be called 'holons'.

The set of primitives idempotents satisfying the relations (1) that have been indexed by using a single symbol and thus we can represent the idempotent by a set of points labelled by a set of integers $i, j.....$ But we also require a way of relating generalize points, i.e. how do we relate $i \to j$, etc. To meet this requirement we will introduce a set of elements $\varepsilon_{ij}$ to denote the relation between i and j. If we impose the well-known multiplication rule

$$\varepsilon_{ik}\varepsilon_{jm} = \delta_{kj}\varepsilon_{im} \qquad (2)$$

then the elements $\varepsilon_{ij}$ will satisfy the relations

$$\varepsilon_{ii}\varepsilon_{ii} = \varepsilon_{ii}$$
$$\varepsilon_{ii}\varepsilon_{jj} = 0 \quad \text{if} \quad ii \neq jj$$

We will label our primitive idempotents with a double index. These elements are then our basic q-numbers.

The discussion so far has been quite general and although we have been using the idea of 'generalized point' there is nothing yet in the structure to indicate that these 'points' can be related to the points of space. In fact, the primitive idempotents are not even unique since it is generally possible to find another set by using some inner automorphism of the algebra. In order to illustrate how the connection with space is made, it is necessary to turn to consider a particular algebraic structure and analyse it in the way we have suggested in this section.

## 3. Definition and structure of the finite Weyl algebra of order n²

We begin by defining the finite Weyl algebra $C_2^n$ of order n² as the polynomial algebra generated over the complex field by the set of generating elements $\{e_0^1, e_1^0\}$ [3] subject to the relations [4]

$$(e_0^1)^n \stackrel{\Delta}{=} e_0^n = 1 \qquad (4)$$

$$(e_1^0)^n \stackrel{\Delta}{=} e_n^0 = 1 \qquad (5)$$

$$e_0^1 e_1^0 = \omega \, e_1^0 e_0^1 \qquad (6)$$

Where $\omega = \exp\{\dfrac{2\pi i}{n}\}$

These relations define the $C_2^n$ algebra completely. Then the element $e_b^a$ takes the form:

$$e_b^a = e_0^a e_b^0 = \omega^{ab} e_b^0 e_0^a$$

We may obtain the general rule of combination:

$$e_b^a e_d^c = \omega^{-bc} e_{b+d}^{a+c} \qquad (7)$$

With this rule we may accomplish all the manipulations of the algebra.

It is easy to demonstrate that the n² elements $e_b^a$ form a basis for the $C_2^n$ algebra. Thus every element of the algebra may be written as

$$A = \sum_{a,b=0}^{n-1} A_{ab} e_b^a \qquad (8)$$

In terms of this basis it is possible to obtain a complete set of pairwise orthogonal primitive idempotents $\varepsilon_{ii}$ one such that that we will use has the form:

$$\varepsilon_{ii} = \frac{1}{n}\sum_k \omega^{-ik} e_k^0 \qquad (9)$$

And satisfies

$$\sum_i \varepsilon_{ii} = 1 \qquad (9a)$$

It is quite easy to see by direct algebraic multiplication and the use of rule (7) that the $\varepsilon_{ii}$ satisfy relations (2) and (3).

One possible set of $\varepsilon_{ij}$ associated with the primitive idempotent given by (9) can be written as

$$\varepsilon_{ij} = \frac{1}{n}\sum_r \omega^{-jr} e_r^{j-i} \qquad (10)$$

Direct multiplication shows that this expression satisfies the multiplication rule (2) viz:

$$\varepsilon_{ik}\varepsilon_{jm} = \delta_{kj}\varepsilon_{im} \qquad (2)$$

Each of the n idempotents defines an dimensional subspace in the $n^2$ dimensional space associated with the $C_2^n$ algebra. To illustrate the method let us arbitrarily choose right and left ideals associated with the idempotent given by the index i = 0

$$\varepsilon_{00} = \frac{1}{n}\sum_k e_k^0 \qquad (11)$$

By employing standard algebraic techniques we can find the right and left ideals $I_R^{(0)}$ and $I_L^{(0)}$ associated with this fundamental idempotent. These ideals form n-dimensional vector subspaces with basis vectors given by

$$I_L^{(0)}(i) = \frac{1}{n}\sum_k e_k^{-i} \qquad (12)$$

$$I_R^{(0)} = \frac{1}{n}\sum_k \omega^{ik} e_{-k}^i \qquad (13)$$

It is now straightforward to show that

$$I_L^{(0)}(i) I_R^{(0)}(j) = \varepsilon_{ij} \qquad (14)$$

and

$$I_R^{(0)}(i) I_L^{(0)}(j) = \delta_{ij}\varepsilon_{00} \qquad (15)$$

Comparison of (12) and (13) with (10) shows us that

$$I_L^{(0)}(i) = \varepsilon_{i0} \text{ and } I_R^{(0)}(j) = \varepsilon_{0j}$$

So that (14) and (15) are consequences of (2).

We may now introduce an operator which, for reasons that will become apparent later, we denote by X. This operator will label the generalized points if we define it as

$$X = \frac{1}{n}\sum_{jk} j\omega^{-jk} e_k^0 = j\varepsilon_{jj} \qquad (16)$$

Then immediately we see

$$X\varepsilon_{jj} = j\varepsilon_{jj} \qquad (17)$$

And

$$X\varepsilon_{jm} = j\varepsilon_{jj} \qquad (18)$$

In this way we see that the decomposition of the identity in terms of the idempotents given by equation (9) provides an order for the set of primitive idempotents and hence an order for the generalized points. But, of course, it is always possible to obtain a new set of primitive idempotents under the inner automorphism

$$\varepsilon'_{jj} = S\varepsilon_{jj}S^{-1} \qquad (19)$$

Where S is any element of $C_2^n$. The new primitive idempotents will also provide an order for the generalized points but, in general, this order will not be simply related to the original order of the generalized points given by the first set of primitive idempotents. Closer examination of the effects of equation (19) suggests a kind of 'exploding' transformation in which each generalized point of the old set is spread out into some or all of the points of the new set. Yet the implications of the notation used in equations(17) and (18) are that X will be a position operator that can be used to locate or label a particular generalized point through an eigenvalues j that reflects the position of the point. However, this order seems to be arbitrary. In section 6 we shall enquire into how one can obtain an invariant meaning to the order rather than just imposing it from the outside.

**4. The geometrical interpretation of the algebra $C_2^n$**

In the usual Cartesian view, we can order a discrete set of equally spaced points on the real line by choosing an origin and defining a unit displacement. Successive applications of this unit displacement will take us through the series of points in the right order. Similarly, in terms of our generalized points, we can choose a basic primitive idempotent $\varepsilon_{00}$ to serve as an origin and select an element $T$ of the algebra to define a unit displacement through the relation

$$\varepsilon_{j+1j+1} = T\varepsilon_{jj}T^{-1} \qquad (20)$$

For the primitive idempotent defined in (11), $T$ Has a very simple form, namely, $T^{-1} = e_0^1$. Using this to define the canonical order, we find

$$\varepsilon_{jj} = e_0^{-j}\varepsilon_{00}e_0^{j} \qquad (21)$$

Since the $\varepsilon_{jj}$ are the generalized points of our structure which are labelled through the eigenvalue equation (17), we can interpret $T = e_0^{-1}$ as a translation operator on our discrete space. Indeed it was this fact that led us to the particular form of the basic idempotent defined in equation (11).

Since the algebra is symmetrical in $e_0^1$ and $e_1^0$ we could raise the question as to whether the other generator $e_1^0$ could be used as a translation operator based on another set of generalized points $\varepsilon_{jj}^{'}$ with

$$\varepsilon_{00}^{'} = \frac{1}{n}\sum_{k}\omega^{-ik}e_0^{k} \qquad (22)$$

So that

$$\varepsilon_{jj}^{'} = e_j^0\varepsilon_{00}^{'}e_{-j}^0 \qquad (23)$$

The generalized points defined by $\varepsilon_{jj}^{'}$ can then be labelled through

$$X^{'}\varepsilon_{jj}^{'} = j\varepsilon_{jj}^{'} \qquad (24)$$

And

$$X^{'}\varepsilon_{jm}^{'} = j\varepsilon_{jm}^{'} \qquad (25)$$

With

$$X^{'} = \frac{1}{n}\sum_{jk}j\omega^{-jk}e_0^{k} \qquad (26)$$

In this discrete space the translation operator is $T^{'} = e_1^0$. Thus we have distinguished two discrete spaces, each comprising a set of points for which the generators define a translation operator via (21) and (23). These spaces are related through the transformation

$$Z\varepsilon_{jj}Z^{-1} = \varepsilon_{jj}^{'} \qquad (27)$$

Where

$$Z = \frac{1}{\sqrt{n^3}}\sum_{ijk}\omega^{j(i-k)}e_k^{j-i} \qquad (28)$$

Thus the symmetry between $e_0^1$ and $e_1^0$ establishes a kind of duality between the two discrete subspaces. Specifically, the points of each space are related via (27) and (28). As we have seen, under any inner-automorphism the new primitive idempotents have a far from simple relationship to the original idempotents, nevertheless there remains a natural complementarity between the two sets of new primitive idempotents and hence between the two discrete spaces constructed from these idempotents.

We are familiar with this kind of natural duality in standard quantum mechanics where translations in space are generated by the momentum operator and translations in momentum space are generated by the position operator. In fact, the operator corresponding to a translation through a distance $a$ is given by

$$T_X(a) = e^{-iaP}$$

This suggests that for a discrete space there exists a momentum operator $P$ such that

$$e_0^1 = e^{\frac{2\pi i P}{n}} \qquad (29)$$

We introduce the factor $\frac{2\pi}{n}$ for later convenience.

Again in standard quantum mechanics a translation operator in momentum space is

$$T_P(a) = e^{iaX}$$

Which again suggests

$$e_1^0 = e^{\frac{-2\pi i X}{n}} \qquad (30)$$

Thus the two discrete spaces generated above are the discrete position and momentum spaces which imply that in equation (26) we should write $X^{'} = P$. In order to confirm this identification let us investigate the commutation between $X$ and $P$ as we go to the limit of $n \to \infty$. To do this let us first form the commutator of X defined by (16) and $P = (X^{'})$ defined by (26).

$$[X,P] = \frac{1}{n}\sum_{jkrs}(s-j)r\omega^{r(s-j)}\omega^{-jk}e_k^{j-s} \qquad (31)$$

Now let us see what happens. To do this, the discrete indices are replaced by continuous indices viz:

$$\frac{1}{n}\sum_{jkrs} \to \frac{1}{2\pi}\iiiint djdkdrds$$

And

$$\omega^\alpha \to \exp\{+2\pi i\alpha\}$$

We have from (31)

$$[X,P] = \frac{1}{2\pi}\iiiint r\exp\{+2\pi i r(s-j)\}(s-j)\omega^{-jk}e_k^{j-s}djdkdrds$$

$$+\frac{1}{i(2\pi)^2}\iiiint \frac{d}{d(s-j)}\exp\{+2\pi i r(s-j)\}(s-j)e_k^{j-s}\exp\{+2\pi ijk\}djdkdrds$$

Integrating over dr

$$=\frac{-i}{2\pi}\iiint \delta'(s-j).(s-j).\exp\{2\pi ijk\}e_k^{j-s}djdkds$$

Integrating over ds

$$[X,P] = i\iint \exp\{2\pi i.jk\}e_k^0 dkdj \qquad (32)$$

Where the latter double integral is the algebraic expression of the unity element 1 since it is none other than the completeness relation for the infinite dimensional case. Therefore

$$[X,P] = i$$

Thus, under the appropriate limiting procedure, the commutator of the $X$ and $P$ elements assumes the quantum mechanical value, confirming our interpretation of both $X$ and $P$.

This result it's not new. It was obtained by Weyl (1932) and more directly by Santhanam (1977), but the method we have used is new and our whole approach throws a different light on quantum mechanics. It is not necessary to assume an a priori externally imposed position order together with an independent and externally imposed momentum order. The appropriate algebra already carries the order of space implicitly, providing the momentum is also part of the same structure. Thus the correlation between $X$ and $P$ has little to do with a duality of 'waves' or 'particles' but has to do with the description of structure process that does not require the external imposition of independent space and momentum orders.

The commutator (32) leads directly to the uncertainty principle and it is natural to assume that the discrete algebra ($n \neq \infty$) should also contain some uncertainty. Santhanam and Tekumalla (1975) state that the discrete case does not have an uncertainty principle; however, we believe this conclusion to be incorrect, as we demonstrate in the following way. It is well known that provided $A$ and $B$ are two hermitian operators then

$$\Delta A \Delta B \geq \frac{1}{2}\langle\|[A,B]\|\rangle$$

In our case $A = X$ and $B = P$. It is easy to demonstrate that $X$ and $P$ are both hermitian and since they have a non-zero commutator (32) there must exist an uncertainty relation between the two observables.

**5 The connection with the Bra-Ket notation**

In order to bring the discussion onto even more familiar ground, let us connect our approach with the usual bra-ket notation. As pointed out by Frescura and Hiley (1980a) there is a very close correspondence between the ket (bra) and the minimum left (right) ideals. Indeed we can write

$$I_L^{(0)}(i) = |i\rangle \quad ; \quad I_R^{(0)}(i) = \langle i|$$

Then (14) and (15) become

$$I_L^{(0)}(i) I_R^{(0)}(j) = \varepsilon_{ij} = |i\rangle\langle j| \qquad (33)$$

And

$$I_R^{(0)}(i) I_L^{(0)}(j) = \delta_{ij}\varepsilon_{00} = \langle i|j\rangle \varepsilon_{00} \qquad (34)$$

Here the label (0) indicates the choice of the basic primitive idempotent. The usual bra-ket notation however suppresses the dependence of the ket (bra) on the choice of the basic primitive idempotent by writing in general

$$I_L^{(m)}(i) I_R^{(m)}(j) = |i\rangle\langle j|$$

And

$$I_R^{(m)}(i) I_L^{(m)}(j) = \delta_{ij}\varepsilon_{mm} = \langle i|j\rangle \varepsilon_{mm}$$

M being an arbitrary choice of basic idempotent. So the bra-ket notation does not exploit all the q-number structure, and works with only one ideal.

Thus we see that when quantum mechanics is viewed from the q-number theory, the ket notation hides the fact that each ket represents an object with two labels. Suppressing this dependence on two labels means that the eigenvalue equation (18) can replace equation (17) so that each point can now be labelled by a ket $|j\rangle$. Then by using (21) we can write

$$|j+a\rangle = e_0^{-a}|j\rangle \qquad (35)$$

Which shows that $e_0^{-a}$ corresponds to a translation operator that takes you from the point labelled by $|j\rangle$ to the point labelled by $|j+a\rangle$. As we know that in the limit as $n \to \infty$ this particular discrete structure becomes continuous and $C_2^n$ approaches the Heisenberg algebra for a one-dimensional continuum in this limit we may write $|j\rangle \to |x_j\rangle$ in which case equation (18) becomes

$$X|x_j\rangle = j|x_j\rangle \qquad (36)$$

In this equation $X$ acts as a position operator on the points representing the individual elements of a given minimal left ideal. It is this structure that forms the starting point of Santhanam's work (1977).

To complete the picture we can introduce kets in the momentum space so that

$$|P_{j+b}\rangle = e_b^0|P_j\rangle$$

Using the primitive idempotents defined in equation (22) it is now easy to construct explicitly the kets in momentum space, Viz

$$|P_j\rangle = \frac{1}{\sqrt{n^3}} \sum_{ik} \omega^{ij} e_k^{-i} \qquad (37)$$

With the relationship

$$|P_j\rangle = \frac{1}{\sqrt{n}} \sum_i \omega^{ij} |x_i\rangle \qquad (38)$$

Which corresponds to the continuum limit

$$|p\rangle = \frac{1}{\sqrt{2\pi}} \int_{-\infty}^{\infty} e^{ipx} |x\rangle dx \qquad (39)$$

Thus (38) represents the finite dimensional version of the Fourier transform which forms the starting point of the recent work of Gudder and Naroditsky (1981), though here, we see that it arises directly from the q-number structure.

The use of equations (25) and (26) with $X' = P$ gives

$$P|p_j\rangle = j|p_j\rangle$$

## 6. Geometric order as emerging from the algebraic structure

Let us now return to the question of how an algebraic or q-number theory may determine its own space structure. We have seen how the set of primitive idempotents used in Section 3 orders the generalized points and that this order is arbitrary. We shall now discuss the question of how this set of idempotents with its attendant order of "points" can be provided with a natural order rather than this order being arbitrarily imposed from outside.

This order, which is intended to be <u>the</u> geometric order as we observe and experience it, is actually already present in the basic equations of all the relevant underlying physical structures (e.g. fields and particles). Such basic equations exploit the continuity of the position representation in a fundamental way. Indeed it is the requirement of the continuity and single-valuedness of <u>all</u> physically significant operators (including the wave function which in the algebraic approach is replaced by an element of the left ideal) is needed to give rise to the correct energy levels and transition probabilities.

Under an inner automorphism, these equations will in general cease to be differential with respect to the original variables because, as pointed out in the discussion around equation (19), this kind of transformation "explodes" each point into a distribution spreading throughout the original space. Thus the basic physically significant operations have ceased to be "local" (in the sense of being constituted out of functions such as $x$ and $\frac{\partial}{\partial x}$, which have zero matrix elements for points that have finite separation).

So far, it has been possible to write the requirement of the continuity and single-valuedness only in the position representation. In other words these conditions are representation dependent. Without these requirements, the theory is physically indeterminate. But in the q-number approach, this representation is arbitrary so that the physical content of quantum mechanics can <u>not</u> be specified in an arbitrary Heisenberg representation (for example, the wave function is not continuous in momentum space). Therefore, without a further specification of what the "position representation" is, the physical consequences of the theory are not defined.

This implies that in a certain sense, the Heisenberg and Schrödinger pictures are not completely equivalent after an "exploding" transformation, a factor that has not generally been properly appreciated. In order to complete the algebraic (or q-number) structure additional new concepts are needed so as to allow us to assert the complete equivalence of the Heisenberg and Schrödinger pictures (i.e., in a way that determines <u>all</u> physical properties, independently of the representation).

In the discrete case, which we are treating in this paper, there are, of course no differential operators. Instead, however, we may introduce <u>neighbourhood operators</u>, from which can be obtained operators that approach differential operators, in the limit as the density of points becomes infinite.

In order to carry through the details we first point out that two different but related approaches to the notion of a generalized point has been used in the previous sections. In Sections 3 and 4 we have used primitive idempotents themselves to represent each point, whereas in Section 5 we showed that we can also associate each basic element of a minimal ideal with an ordered set of points in configuration space. Although this involves suppressing some of the information carried by the q-numbers, it does enable us to connect directly with the ket formalism where each set of kets $|x_j\rangle$ becomes associated with a one-dimensional discrete space. It is this approach that enables us to bring out more completely the connection in the limit between the neighbourhood operators and differential operators.

We can bring out what is meant by replacing the differential operator of a typical physical equation (e.g., a wave equation) by difference operators, involving neighbouring points in a discrete array. Thus consider replacing the continuous field functions $\psi(x)$ with a corresponding set of discrete values $\psi_j = \langle x_j | \psi_j \rangle$ over our array of points for the index $j$. We then replace

$$\frac{\partial \psi}{\partial x} \to \psi_{j+1} - \psi_{j-1}$$

And

$$\frac{\partial^2 \psi}{\partial x^2} \to \psi_{j+1} - 2\psi_j + \psi_{j-1}$$

Then a typical wave equation would take the form

$$\frac{\partial^2 \psi}{\partial t^2} = \alpha(\psi_{j+1} - 2\psi_j + \psi_{j-1})$$

Where $\alpha$ is a suitable constant.

Under an inner automorphism the point represented by $|x_j\rangle$ will go over into a linear combination

$$|x_j\rangle = \sum_k C_{jk}|x_k\rangle$$

So that

$$\psi_j = \sum_k C_{jk}^* \psi_k$$

The combination $\psi_{j+1} - \psi_{j-1}$ goes into $\sum_k (C_{k,j+1}^* - C_{k,j-1}^*)\psi_k$. It is clear that the difference operator has been replaced by another one which is "non-local", i.e. one that connects points from all over the space (as if each point had been "exploded" into the whole space). This is an example of how the form of the equation has been radically altered by an inner automorphism.

To discuss the implications of this fact further, we define two neighbourhood operators, $N^+$ and $N^-$, which are determined by the equations

$$N^+ \psi_j = \psi_{j+1}$$

(40)

$$N^- \psi_j = \psi_{j-1}$$

Under a general automorphism, $\psi_j = \sum_k C_{jk}\phi_k$, these operators become

$$(N^+)\phi_k = \sum_l \sum_j C_{jk}^* C_{j+1,l}\phi_l$$

(41)

$$(N^-)\phi_k = \sum_l \sum_j C_{jk}^* C_{j-1,l}\phi_l$$

Hence the neighbourhood operators no longer have the simple intuitive meaning they had in equation (40), and in the limit of an infinitely dense array continuity has lost its simple meaning. But as we shall see, even though the "space" itself loses its "local" properties under such a transformation, the physical content of the theory i.e. as reflected in the measurable quantities, is invariant. Indeed we have something analogous to a hologram in which the locality relations are carried non-locally throughout the hologram. Non-locality is thus implicitly incorporated into quantum mechanics and is a direct result of associating position space with a basis not invariant to similarity transformation. Thus, within the algebraic and quantum mechanical formalisms, the physical relationships (which are independent of a change of basis) are no longer tied to the position space stage upon which those events are acted out. In this way, we have been led to question the need for an a priori given space-time manifold. Indeed, this approach points to a deeper structure underlying the space-time manifold, with its own, more general, invariance features.

Clearly, however, in order to define the geometric content of the present theory we need to specify some form of "locality relations". For suppose that a certain structure of algebraic equations in a general Heisenberg representation is given, then we have to add a set of neighbourhood operators and if these reduce to the canonical form (40), a set of "positions" will be defined uniquely in their proper order. Furthermore, if the physical equations are all "local" in the sense defined above, they will involve only these neighbourhood operators. The arbitrariness of what is to be meant by the proper "position representation" will thus be removed, and the physical significance of the algebraic structure will thus remain invariant.

A closer examination of the equation (40) together with the discussions in Section 4 shows that our neighbourhood operators are simply related to the formal elements $e_0^1$ and $e_0^{-1}$ of the Weyl algebra. In consequence the momentum P can be obtained as a suitable limit of $(N^+ - N^-)$ from which it follows that the momentum is in essence the usual displacement to the next nearest neighbour. From the above, it follows that we have not introduced any new content into the theory that was not already contained in the original representation dependent theory. We have mainly confined ourselves to emphasising the new key role played by the neighbourhood relations.

Let us now turn to discuss the possibility of adding new content. Consider first the analogy of the metric in general relativity. The canonical diagonal metric $g_{\mu\mu}$ can be turned by a coordinate transformation $x^\mu = \phi^\mu(x^\nu)$, into

$$\sum_\mu \frac{\partial \phi^\mu}{\partial x^\rho} \frac{\partial \phi^\mu}{\partial x^\sigma} g_{\mu\mu} = g'_{\rho\sigma}$$

Now as long as a global transformation $\phi^\mu(x^\nu)$ exists there is no new content in this theory. However when the curvature is nonzero only local transformations can be found, which reduce the matrix to diagonal form, and we have the possibility of new content, namely, gravitation. We can apply a similar argument to the neighbourhood matrix. In the present theory it is assumed that the matrix can be reduced everywhere by a unitary transformation. However it may not be possible to make neighbourhood matrices all diagonal all together. Thus we have the possibility of introducing new content.

In summary then, what we have done is in effect to treat the algebraic structure as a pre-space, whose "points" are related to ordinary "points" by an exploding transformation (or equivalently, these "points" may be described as being in an implicate order). We have then shown how a unique geometric (or explicate) order can be determined by means of the neighbourhood operators, which are part of the same algebraic structure. So our ordinary physical space (i.e., the space which describes basic entities such as fields and particles) emerges from the pre-space.

It must be emphasized that the main purpose of this paper has been to draw attention to the possibilities latent in the algebraic structure, and that what has been done is only a beginning of what is needed for a full realization of these possibilities. Firstly, we have to go to the continuous limit, where new physically significant features arise. In particular, one will then require the continuity of all physically significant features, and this is a further restriction, which has no meaning in a discrete array. Then, we will have to bring in time as well as space, and incorporate general as well as special relativity. We are currently working on these questions, which we hope to treat in further publications.

ACKNOWLEDGMENTS

We would like to thank Dr. F.A.M. Frescura for helpful discussions. P.G. Davies would like to thank the S.R.C. for a grant which enabled this work to be completed.

**References**

Birkoff, G. and von Newmann, J. (1936)

Bohm, D (1973). Foundations of Physics, 3, 139

Born, M. and Jordan, P. (1925) Zeitschrift für Physick, 34, 858

Dirac, P.A.M. (1963) Physical Review, 139B, 684.

Dirac, P.A.M. (1974) The Principles of Quantum Mechanics, Oxford.

Eddington, A.S. (1946) Fundamental Theory, Cambridge.

Finkelstein, D. (1972) Physical Review, 5D, 320.

Frescura, F.A.M. and Hiley, B.J. (1980a) Foundations of Physics, 10, 7.

Frescura, F.A.M. and Hiley, B.J. (1980b) Foundations of Physics, 10, 705.

Gudder, S. and Naroditsky, V. (1981) International Journal of Theoretical Physics, 20, 619

Hawking, S.W. (1978) Nuclear Physics, B144, 349.

Hiley, B.J. (1980) Annal de la Foundation Louis de Broglie 5, 75

Jauch, J.M. (1968). Foundations of Quantum Mechanics, Addison Wesley, Massachusetts.

Misner, C.W., Thorne, K. and Wheeler, J.A. (1973) Gravitation, Freeman, San Francisco.

Morris, A.O. (1967) Quarterly Journal of Mathematics, Oxford. (2), 18, 7.

Patton, C.M. and Wheeler, J.A. (1975) Quantum Gravity, edited by Islam, Penrose and Sciama, Oxford.

Reisz, M (1946) Comptes Rendus 10$^{me}$ Congrès Mathematique

Sakharov, A.D. (1967) Doklady Akademy Nauk, U.S.S.R. 177, 70 (English translation, Soviet Physics Doklady, 12, 1040 (1981)).

Santhanam, T.S. and Tekumalla, A.R. (1975) Foundations of Physics, 6, 583.

Santhanam, T.S. (1977) In the Uncertainty Principle and Foundations of Quantum Mechanics, edited by Price and Chissick, Wiley, London

t'Hooft, G. (1978) Recent Developments in Gravitation. Plenum Press.

Taylor, J.G. (1979) Physical Review, 19D, 2336

Weyl, H., (1932) The Theory of Groups and Quantum Mechanics, Reprinted by Dover Publications Inc., New York, USA (1950)

Wheeler, J.A. (1980) Quantum Theory and Gravitation, Ed. Marlow, Academic Press.

---

[1] We realize that Dirac's q-numbers are more general than the q-numbers considered in this paper. Nevertheless we hope this limited investigation will help throw some light on the more general problem.

[2] Greater detailed may be found in chapter 44 of Misner, Thorn and Wheeler 1973, Patton and Wheeler 1975 and also in Wheeler 1980. In presenting this account we are merely tracing the historic evolution of the ideas without commitment to the actual concepts used in the development.

[3] The $\{e_0^1, e_1^0\}$ will be related to the generators of the translation operators in the position and momentum space in the limit as n → ∞ (see section 4)

[4] The symbol $\stackrel{\Delta}{=}$ means it is a definition.

---